\documentclass[aps,prc,twocolumn,showpacs,groupedaddress,floatfix]{revtex4}

\usepackage{graphicx}
\usepackage{dcolumn}
\usepackage{amsfonts}

\begin{document}

\title{Study of $^{64}$Ni+$^{132}$Sn Fusion with Density Constrained TDHF Formalism}

\author{A.S. Umar and V.E. Oberacker}
\affiliation{Department of Physics and Astronomy, Vanderbilt University,
             Nashville, Tennessee 37235, USA}

\date{\today}


\begin{abstract}
We study fusion reactions of the $^{64}$Ni+$^{132}$Sn system using the recently introduced
density constrained time-dependent Hartree-Fock formalism. In this formalism the
fusion barriers are directly obtained from TDHF dynamics. In addition, we incorporate the
entrance channel alignment of the deformed (oblate) $^{64}$Ni nucleus due to dynamical Coulomb excitation.
We discuss the influence of particle transfer and other dynamical effects on the fusion cross sections.
Calculated cross sections are in very good agreement with data and other calculations.
\end{abstract}
\maketitle


\section{\label{sec:intro}Introduction}
With the increasing availability of radioactive ion-beams~\cite{DOE02} the study of
structure and reactions of exotic nuclei has become feasible~\cite{Ji04}.
In particular, detailed investigations of the fusion process are crucial for the prediction
of heavy-element formation and will lead to a better understanding of the interplay among the strong, Coulomb,
and weak interactions as well as the enhanced correlations present in these many-body
systems.

Recently, particular attention has been given to the $^{64}$Ni+$^{132}$Sn system,
where a large sub-barrier fusion enhancement was observed~\cite{Li03,Li05}.
For this system fusion cross sections were measured in the energy range 142~MeV~$\le E_{\mathrm{c.m.}} \le$~195~MeV.
In particular, it was found that fission is negligible for $E_{\mathrm{c.m.}}\le$~160~MeV
and therefore the evaporation residue cross sections have been taken as fusion cross sections.
The enhancement of sub-barrier fusion was originally deduced from comparison with a barrier penetration
calculation, using a phenomenological Woods-Saxon interaction potential whose parameters were
fitted to reproduce the evaporation residue cross sections
for the $^{64}$Ni+$^{124}$Sn system~\cite{Li03,Li05}.
Similarly, early coupled-channel calculations,
which are known to enhance the fusion cross sections by considering coupling to various excitation
channels and neutron transfer, have significantly underestimated the low energy fusion cross
sections for the $^{64}$Ni+$^{132}$Sn system~\cite{Li03}. Subsequently, more sophisticated coupled-channel
calculations lead to an improvement for the description of the lower energy data. Finally, the
inclusion of the neutron transfer channels with positive $Q$ value in addition to inelastic excitations
resulted in the best description to date~\cite{Li07}.

The theoretical analysis of the fusion data generally involves determination
of a phenomenological ion-ion potential such as the Bass model~\cite{Ba74,Ba80},
the proximity potential~\cite{BR77,RV78,SG84,BH78}, or potentials obtained via
the double-folding method~\cite{SL79,BB77,RO83a,RO83b}. Subsequently, the actual
fusion cross section is calculated by either using barrier penetration
models~\cite{Ba80,TB84,RO83a,BT98}, or the coupled-channel method~\cite{LP84,RP84,HR99,Esb04,Esb05}.
The latter includes various excitations of the
target and/or projectile using the coupled-channel formalism~\cite{HR99,Esb04}, as well as
the inclusion of neutron transfer, and can be consistently applied at energies above
and below the barrier~\cite{BT98}. Effectively, the inclusion of each additional excitation
leads to a modification of the original inert core ion-ion potential, resulting in a series
of effective barriers.
One common physical assumption used in many of these
calculations is the use of the frozen density or the sudden
approximation. In this approximation the nuclear densities
are unchanged during the computation of the ion-ion potential as a function
of the internuclear distance. Furthermore, the effects included in channel couplings
are usually based on the static properties of the participating nuclei, which may
accurately represent the early stages of the collision process, but are expected
to change as the two ions strongly interact. While these methods provide a useful
and productive means for quantifying multitudinous reaction data it is desirable to
include dynamical effects and make contact with the microscopic theories of nuclear
structure and reactions.

Recently, we have developed a new approach for calculating heavy-ion interaction
potentials which incorporates all of the dynamical entrance channel effects included in the
time-dependent Hartree-Fock (TDHF) description of the collision process~\cite{UO06b}.
These effects include the neck formation, particle exchange, internal excitations,
and deformation effects to all order, as well as the effect of nuclear alignment
for deformed systems. The method is based on the TDHF
evolution of the nuclear system coupled with density-constrained Hartree-Fock
calculations to obtain the ion-ion interaction potential. Preliminary calculations
for the $^{64}$Ni+$^{132}$Sn system highlighted the importance of dynamical
deformation effects~\cite{UO06d}. Here we give a completed study of fusion cross
sections using this formalism.

In the next section we will summarize some theoretical aspects of the density constrained
TDHF theory along with methods to incorporate dynamical alignment into our calculations,
as well as the method used the calculate cross sections from the resulting barriers.
In Section~\ref{sec:results}
we present interesting aspects of the reaction dynamics and compare our results with
experiment and other calculations.

\section{\label{sec:formal}Theoretical Methods}

\subsection{\label{sec:dctdhf} Density constrained TDHF method}
In this subsection we give a qualitative description of the density constrained
TDHF method which is used to obtain the dynamical barriers for the $^{64}$Ni+$^{132}$Sn
system. Further details of the method can be found in Ref.~\cite{UO06b}.

The {\it density constraint} is a novel numerical method that was developed in
the mid 1980's~\cite{CR85,US85} and was used to provide a microscopic
description of the formation of shape resonances in light systems~\cite{US85}.
In this approach the TDHF time-evolution takes place with no restrictions.
At certain times during the evolution the instantaneous density is used to
perform a static Hartree-Fock minimization while holding the total density constrained
to be the instantaneous TDHF density. In essence, this provides us with the
TDHF dynamical path in relation to the multi-dimensional static energy surface
of the combined nuclear system. Since we are constraining the total density
all moments are simultaneously constrained.
The numerical procedure for implementing this constraint and the method for
steering the solution to $\rho_{\mathrm{TDHF}}(\mathbf{r},t)$ is discussed in Refs.~\cite{CR85,US85}.
The convergence property is as good if not better than in the traditional
constrained Hartree-Fock
calculations with a constraint on a single collective degree of freedom.

In Ref.~\cite{UO06b} we have shown that the ion-ion interaction potential is simply
given by
\begin{equation}
V(R)=E_{\mathrm{DC}}(R)-E_{\mathrm{A_{1}}}-E_{\mathrm{A_{2}}}\;,
\label{eq:vr}
\end{equation}
where $E_{\mathrm{DC}}$ is the density constrained energy at the instantaneous
separation $R(t)$, while $E_{\mathrm{A_{1}}}$ and $E_{\mathrm{A_{2}}}$ are the binding energies of
the two nuclei obtained with the same effective interaction. We would like to
emphasize again that this procedure does not affect the TDHF time-evolution and
contains no {\it free parameters} or {\it normalization}. In practice,
TDHF runs are initialized with energies above the Coulomb barrier and in
Ref.~\cite{UO06b} we have shown that there is no appreciable energy dependence
to the barriers obtained via the density constrained TDHF method.
A detailed description of our new three-dimensional unrestricted TDHF code
has recently been published in Ref.~\cite{UO06}.
For the effective interaction we have used the Skyrme SLy5 force~\cite{CB98}
including all of the time-odd terms.

\subsection{\label{sec:align} Fusion with alignment}
In general, the fusion cross sections depend on the interaction potential and form factors in the
vicinity of the Coulomb barrier. These are expected to be modified during the collision due
to dynamical effects. In addition, experiments on sub-barrier
fusion have demonstrated a strong dependence of the total fusion cross section
on nuclear deformation~\cite{SE78}. The dependence on nuclear orientation has received
particular attention for the formation of heavy and superheavy elements~\cite{KH02}
and various entrance channel models have been developed to predict its role in
enhancing or diminishing the probability for fusion~\cite{SC04,UO06c}.
Recently, we have developed a new approach for calculating the effect of nuclear alignment
for deformed systems~\cite{UO06c}. In essence, the procedure for incorporating alignment
into the
evolution of the heavy-ion collision dynamics is done in two separate steps:
a) A dynamical Coulomb alignment calculation to determine the probability that a given nuclear orientation
occurs at the distance $R(t_0)$, where the TDHF run is initialized.
The alignment generally results from multiple {\it E2}/{\it E4} Coulomb excitation
of the ground state rotational band.
The distance $R(t_0)$ is chosen such
that the nuclei only interact via the Coulomb interaction.
b) A TDHF calculation, starting at this finite internuclear distance $R(t_0)$, for
a fixed initial orientation of the deformed nucleus.
Since the experiments are usually done with unpolarized beams, in
a full quantum mechanical calculation one would have to average over discrete
quantum mechanical rotational bands. In the classical limit, this
corresponds to averaging over orientation angles. A general study of
taking the classical limit of the relative nuclear motion during a
heavy-ion collision which includes inelastic excitations of one of
the heavy ions in the entrance channel has been given in Ref.~\cite{USE84}.

The heavy-ion interaction potential between two deformed nuclei depends on the distance
vector between their centers-of-mass, ${\bf R}$, and on the relative orientation of their
intrinsic principal axis systems which may be described in terms of three Euler angles
$\alpha,\beta,\gamma$ per nucleus, i.e. in the most general case we have
\begin{equation}
V = V ( {\bf R},\alpha^{(1)},\beta^{(1)},\gamma^{(1)},\alpha^{(2)},\beta^{(2)},\gamma^{(2)} )\;.
\end{equation}
Explicit expressions for this interaction potential within the double-folding
method are given in Ref.~\cite{RO83b}. The expression for $V$ can be simplified
if the intrinsic nuclear density distributions are axially symmetric;
in this case, the potential does not depend on the Euler angles $\gamma^{(1)},\gamma^{(2)}$
which describe rotations about the symmetry axes. If we put, for convenience,
the distance vector in $z$-direction, ${\bf R}= R {\bf e}_z$, the potential
between two deformed axially symmetric nuclei has the structure
\begin{equation}
V = V ( R,\beta^{(1)},\beta^{(2)}, \Delta \alpha)\;.
\end{equation}
Finally, if one of the nuclei is spherical, e.g. nucleus (1),
the potential is simply given by
\begin{equation}
V = V(R,\beta^{(2)})
\end{equation}
where the Euler angle $\beta^{(2)}$ describes the direction of the nuclear symmetry axis
relative to the internuclear distance vector.

\begin{figure}[!htb]
\includegraphics*[scale=0.44]{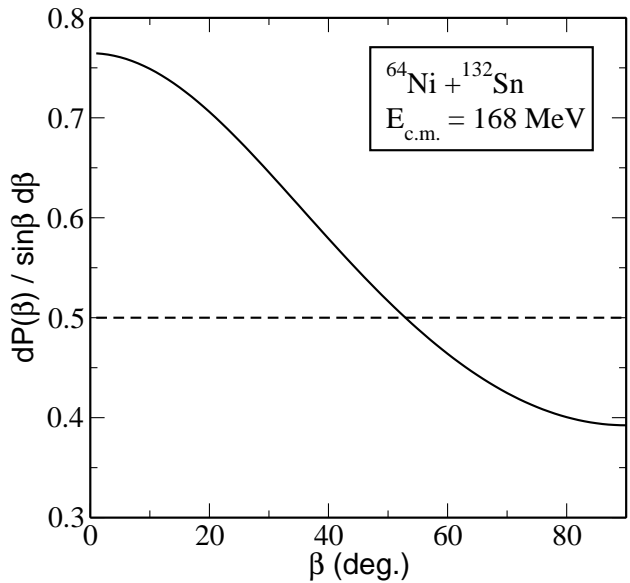}
\caption{\label{fig:align_ni_sn} Dynamic alignment due to Coulomb
excitation of $^{64}$Ni. Shown is the orientation probability
as a function of the Euler angle $\beta$ in a central collision at internuclear
distances $R=1500$~fm (dashed curve) and at $R=16$~fm (solid curve).}
\end{figure}

Details of the dynamic alignment formalism are presented in~\cite{UO06c}. We give here
a brief summary: For a given incident energy $E_{\mathrm{c.m.}}$ we carry out a semiclassical
Coulomb excitation calculation of the dominant collective levels of the deformed
nucleus. The energy levels and $EL$-transition matrix elements for $^{64}$Ni
are taken from experimental data \cite{ENSDF}:  $E_{2+}=1.346$~MeV, $E_{4+}=2.610$~MeV
and $M(E2, 0+ \rightarrow 2+) = -27.0\ e~$fm$^2$  (oblate deformation).
The Coulomb excitation calculation starts at very large internuclear distances
(about $1500$~fm) when both nuclei may be presumed to be in their respective
ground states and stops at the ion-ion separation distance $R(t_0)$ (about $16$~fm). The Coulomb
excitation amplitudes determine the probability distribution of initial orientations.
In Fig.~\ref{fig:align_ni_sn} we show the differential alignment probability
as a function of the Euler angle $\beta$ used in our calculations.

In the case of one spherical nucleus ($^{132}$Sn) and one deformed reaction partner ($^{64}$Ni), the total fusion cross
section is given by an integral over all orientation (Euler) angles, with solid
angle element $d\Omega=2\pi sin\beta d\beta$
\begin{equation}
\label{eq:fusion}
\sigma(E_{\mathrm{c.m.}}) = \int d\Omega\; \frac{dP}{d\Omega}\; \sigma(E_{\mathrm{c.m.}},\Omega)\;,
\end{equation}
where $dP/d\Omega$ represents the alignment probability and $\sigma(E_{\mathrm{c.m.}},\Omega)$
is the fusion cross section associated with a particular alignment.

\subsection{\label{sec:cross} Cross section calculation}
For a consistent calculation of fusion cross sections at above
and below the barrier energies we have adopted the commonly used
{\it incoming wave boundary condition} (IWBC) method~\cite{Raw64,LP84}.
Once the ion-ion potential for a particular orientation, $V(R,\beta)$,
is calculated the two-body Schr\"odinger equation becomes
$$
\left[ \frac{-\hbar^2}{2\mu}\frac{d^2}{dR^2}+\frac{L(L+1)\hbar^2}{2\mu R^2} + V(R,\beta)-E\right] \psi_L(R,\beta)=0\;,
$$
where $\mu$ is the reduced mass and $L$ denotes orbital angular momentum.
IWBC assumes that once the minimum of the potential is reached fusion will
occur, consequently no imaginary part of the potential is needed. In
practice, the Schr\"odinger equation is integrated from the potential
minimum, $R_{min}$, where only an incoming wave is assumed, to a large asymptotic distance,
where it is matched to incoming and outgoing Coulomb wavefunctions to obtain the
penetration factor, $P_L(E,\beta)$, as the ratio of the incoming flux at $R_{min}$
to the incoming Coulomb flux at large distance. The total cross section is given by
\begin{equation}
\sigma(E,\beta)=\frac{\pi}{k_0^2}\sum_L (2L+1)P_L(E,\beta)\;,
\end{equation}
with $k_0=\sqrt{2\mu E}$.
For the numerical implementation we have followed the procedure
for the coupled-channel code CCFUL described in Ref.~\cite{HR99}
and exactly reproduced their results for an inert-core potential.

\section{\label{sec:results} Results}
We have carried out a number of TDHF calculations with accompanying
density constraint calculations to compute $V(R,\beta)$ given by Eq.~(\ref{eq:vr}).
A detailed description of our new three-dimensional unrestricted TDHF code
has recently been published in Ref~\cite{UO06}.
For the effective interaction we have used the Skyrme SLy5 force~\cite{CB98}
including all of the time-odd terms. The code was modified to self-consistently
generate initial states for $^{64}$Ni with different orientations. All of our
TDHF calculations were done at an initial energy of $E_{\mathrm{c.m.}}=168$~MeV
and separation $R(t_0)=16$~fm. As we have reported in Ref.~\cite{UO06b} the
potential barriers obtained from the density constrained TDHF method are not
sensitive to the initial energy (above the barrier). We have
tested this again by running a few orientations at $158$~MeV and $176$~MeV and did
not observe any appreciable difference.

\subsection{\label{sec:dynamic} Particle exchange}
In a TDHF collision leading to fusion the relative kinetic
energy in the entrance channel is entirely converted into internal
excitations of a single well defined compound nucleus. In TDHF theory
the dissipation of the relative kinetic energy into internal excitations is
due to the collisions of the nucleons with the ``walls'' of the
self-consistent mean-field potential. TDHF studies demonstrate that the
randomization of the single-particle motion occurs through repeated
exchange of
nucleons from one nucleus into the other. Consequently, the equilibration of
excitations is very slow, and it is sensitive to the details of the
shape evolution of the composite system.
This is in contrast to
most classical pictures of nuclear fusion, which generally assume near
instantaneous, isotropic equilibration.
\begin{figure}[!htb]
\includegraphics*[scale=0.40]{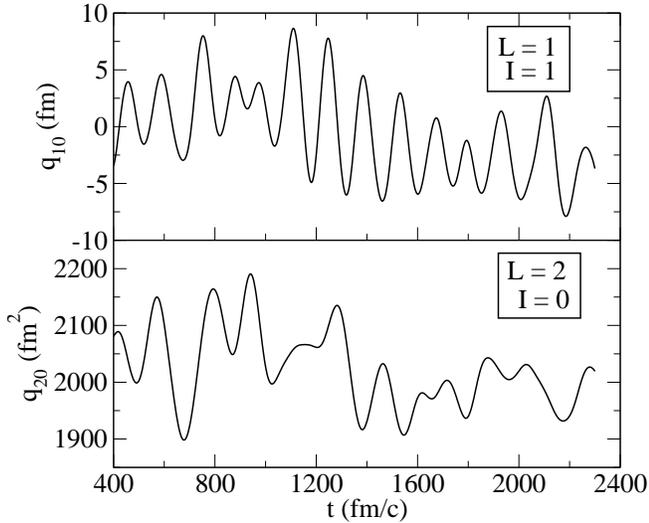}
\caption{\label{fig:qt} Time-dependence of the isovector dipole and isoscalar
quadrupole moments for the head-on collision of $^{64}$Ni+$^{132}$Sn system
at $E_{\mathrm{c.m.}}=168$~MeV and $\beta=90^{\circ}$. }
\end{figure}

Recently, the importance of transfer of neutrons with positive $Q$ value
in fusion has been emphasized~\cite{Li07,ZSW07,DLW83}. In TDHF
this effect manifests itself as the excitation of the pre-compound collective
dipole mode, which is likely when ions have significantly different $N/Z$
ratio, and is a reflection of dynamical charge equilibration. In Fig.~\ref{fig:qt}
we show the time evolution of the isovector dipole and isoscalar
quadrupole moments for the head-on collision of $^{64}$Ni+$^{132}$Sn system
at $E_{\mathrm{c.m.}}=168$~MeV and $\beta=90^{\circ}$. The Fourier transform
of these oscillations show a $9$~MeV isovector dipole peak as well as
$4$~MeV and $7$~MeV isoscalar quadrupole peaks.
\begin{figure}[!htb]
\includegraphics*[scale=0.43]{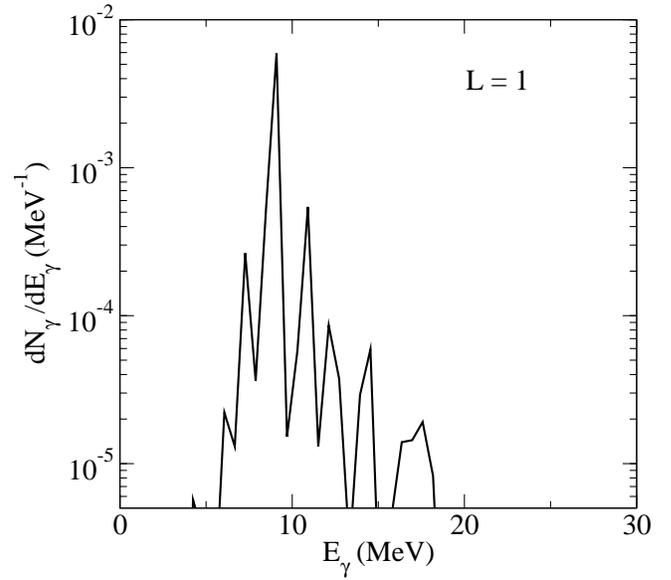}
\caption{\label{fig:yield} Dipole gamma yield for the
head-on collision of $^{64}$Ni+$^{132}$Sn system
at $E_{\mathrm{c.m.}}=168$~MeV and $\beta=90^{\circ}$. }
\end{figure}
For a time-dependent charge distribution it is possible to calculate the
$\gamma$-yield as a function of gamma energy~\cite{US85}, $E_{\gamma}$.
The asymptotic yield integrated over a spherical surface is given by
\begin{equation}
\label{eq:yield}
\frac{dN_{\gamma}}{dE_{\gamma}}=\frac{\hbar c}{8\pi^{2}}
\frac{1}{(\hbar \omega )^{3}}\sum _{L}(2L+1)|a(L,\hbar \omega)|^{2}\;,
\end{equation}
where the amplitudes $a(L,\hbar \omega)$
\begin{equation}
a(L,\hbar \omega )=\frac{4\pi}{i(2L+1)!!}\left[\frac{L+1}{L}\right]^{1/2}
\left[\frac{\hbar \omega}{\hbar c}\right]^{L+2}M_{L}(\hbar \omega )\;,
\end{equation}
are given in terms of the Fourier transform of the moments of the density
\begin{equation}
M_{L}(t)=\int d^{3}rr^{L}Y_{L0}(\hat{r})\rho(\mathbf{r},t)\;.
\end{equation}
Computation of the yield given by Eq.~\ref{eq:yield} shows that the
dominant contribution is from the dipole mode, which is shown in Fig.~\ref{fig:yield}.
\begin{figure}[!htb]
\includegraphics*[scale=0.43]{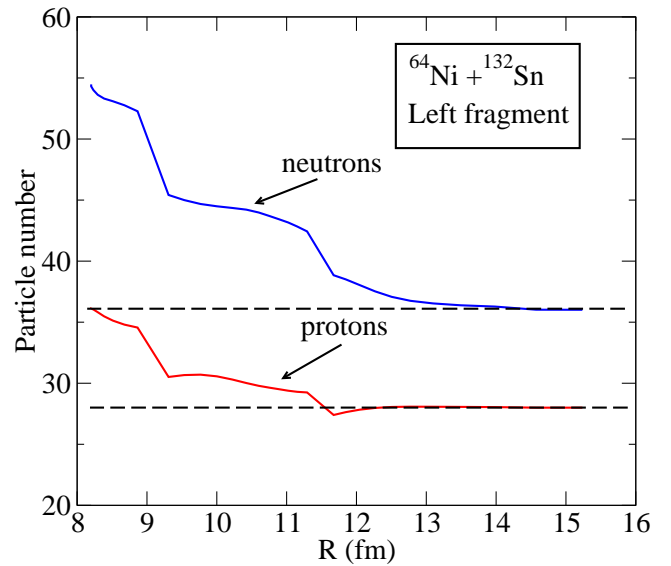}
\caption{\label{fig:left} (Color online) Change in neutron (blue curve) and proton (red curve)
numbers of the left fragment (originally $^{64}$Ni) as a function of the
ion-ion separation coordinate, $R$. The dashed lines are there to emphasize the
asymptotic values.}
\end{figure}

While the above analysis provides a picture of particle exchange dynamics over
time it is also possible to examine particle exchange during the initial stages
of the collision, namely from the well separated nuclei to the time when the
minimum separation is reached. In Fig.~\ref{fig:left} we show the neutron and
proton transfer to the left fragment, which in our case is initially $^{64}$Ni.
As we can see in Fig.~\ref{fig:left} the number of neutrons and protons increase
for the left fragment as the ions come into contact. As expected the transfer of
neutrons starts earlier since they are dominant in the surface region. For small
$R$ values the increase is simply due to charge equilibration as the two nuclei
have a substantial overlap. However, at a separation of $12$~fm, which corresponds
approximately to the top of the potential barrier as discussed in the next subsection,
we have about two neutrons transferred, and at $11$~fm we have as many as six neutrons
transferred. It is also interesting to observe that the large transfers in the
early stages of the collision happen quickly, indicated by the sudden jumps in
neutron number. Naturally, the right fragment undergoes just the opposite of these
trends.

\subsection{\label{sec:pot} Dynamical potentials and mass}
\begin{figure}[!htb]
\includegraphics*[scale=0.42]{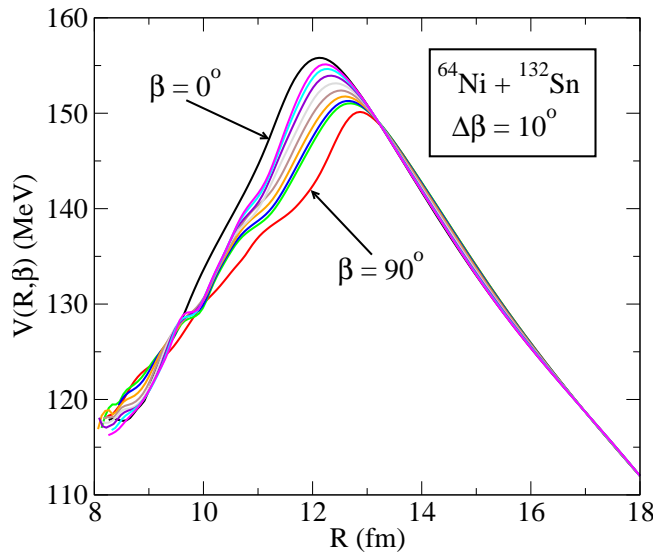}
\caption{\label{fig:vrb} (Color online) Potential barriers, $V(R,\beta)$, obtained from density constrained
TDHF calculations for the $^{64}$Ni+$^{132}$Sn system. Angle $\beta$ indicates different
orientations of the deformed $^{64}$Ni nucleus in $\Delta\beta=10^{\circ}$ intervals.}
\end{figure}

In this subsection we discuss the barriers obtained via the
density constrained TDHF method, and other related quantities,
such as the effective mass and the reduced mass. We have performed
calculations of $V(R,\beta)$ in $\Delta\beta=10^{\circ}$ intervals
from $\beta=0^{\circ}$ to $\beta=90^{\circ}$. In Fig.~\ref{fig:vrb}
we show all of these barriers. The lowest barrier corresponds to
$\beta=90^{\circ}$ orientation of the symmetry axis of the oblate
$^{64}$Ni nucleus with respect to the collision axis. Each subsequent barrier
is reduced by $10^{\circ}$ up to the highest barrier at $\beta=0^{\circ}$
orientation.
For the case of $\beta=0^{\circ}$ orientation the calculated barrier is almost exactly
the same as the empirical barrier used in Ref.~\cite{Li03} without channel couplings,
having a barrier height of $155.8$~MeV and location of the barrier peak at approximately
$R_B=12.1$~fm. The difference for smaller $R$ values is due to the use of the point Coulomb interaction
in the model calculation, which is unphysical when nuclei overlap. As seen in Fig.~\ref{fig:vrb} for
higher $\beta$ values the barrier is lowered and the barrier peak moves to larger $R$ values, with
the lowest barrier having a height of $150.1$~MeV and peaking around $R_B=13$~fm.
The physical picture which emerges from these calculations is that for all energies above $150.1$~MeV,
which is the peak of the lowest barrier, the fusion cross section will be dominated by lower barriers
as the cross section above the barrier is substantially larger than sub-barrier cross sections. In other
words the only experimental data point that appears to be truly sub-barrier is the lowest energy point at $142.6$~MeV.
\begin{figure}[!htb]
\includegraphics*[scale=0.42]{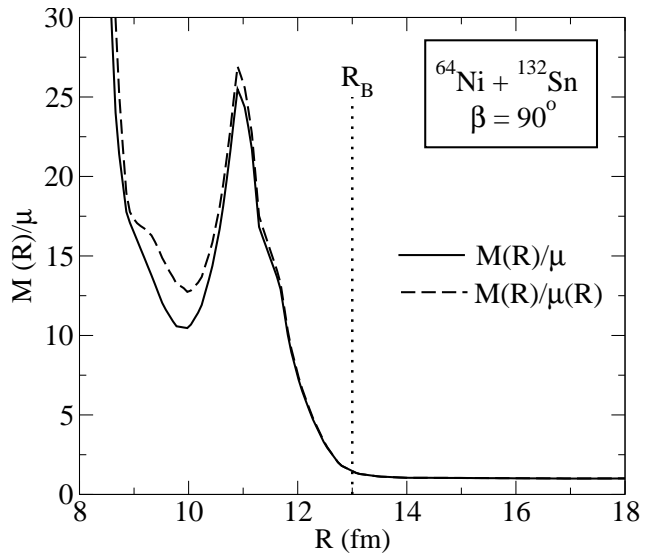}
\caption{\label{fig:eff} Effective mass calculated from Eq.~(\protect\ref{eq:eff})
for the head-on collision of $^{64}$Ni+$^{132}$Sn system at $E_{\mathrm{c.m.}}=168$~MeV and $\beta=90^{\circ}$.}
\end{figure}

In addition to all of the dynamical effects included in the potentials $V(R,\beta)$ it is also
possible to construct effects due to dynamical mass. Since in the density constrained TDHF method the
potential is obtained from the TDHF evolution by essentially extracting the internal excitation
energy from instantaneous TDHF solutions via the density constraint, the energy conservation becomes
\begin{equation}
E_{\mathrm{c.m.}}=\frac{1}{2}M(R,\beta){\dot{R}}^{2}+V(R,\beta)\;.
\end{equation}
For a particular initial orientation at asymptotic energy $E_{\mathrm{c.m.}}$ we obtain
the collective velocity $\dot{R}$ directly from the TDHF evolution and the potential
$V(R,\beta)$ from the density constraint calculations. Thus, the effective mass is given by
\begin{equation}
\label{eq:eff}
M(R,\beta)=\frac{2\left[E_{\mathrm{c.m.}}-V(R,\beta)\right]}{{\dot{R}}^{2}}
\end{equation}
In Fig.~\ref{fig:eff} we show the effective mass as a function of the ion-ion separation
coordinate $R$ for energy $E_{\mathrm{c.m.}}=168$~MeV and $\beta=90^{\circ}$.
The solid curve is the effective mass obtained from Eq.~(\ref{eq:eff}) scaled with the
constant reduced mass $\mu$ for the two ions. The ratio starts from unity at large distances
and increases as $R$ gets smaller and as the ions pass the top of the barrier as indicated by
the dotted line in Fig.~\ref{fig:eff}. This is due to the fact that the relative velocity,
$\dot{R}$, becomes small as the ions begin to overlap substantially, while the numerator of
Eq.~\ref{eq:eff} remains non-zero since the energy is above the barrier. After the initial
slowdown the ions accelerate once more before reaching the composite system for which $\dot{R}=0$.
A similar behavior is observed for all values of $\beta$. Traditionally, the effective mass is obtained
from constrained Hartree-Fock (CHF) or adiabatic time-dependent Hartree-Fock (ATDHF) calculations~\cite{GRR83}
and shows a strikingly similar behavior to the ones obtained through our method.
\begin{figure}[!htb]
\includegraphics*[scale=0.42]{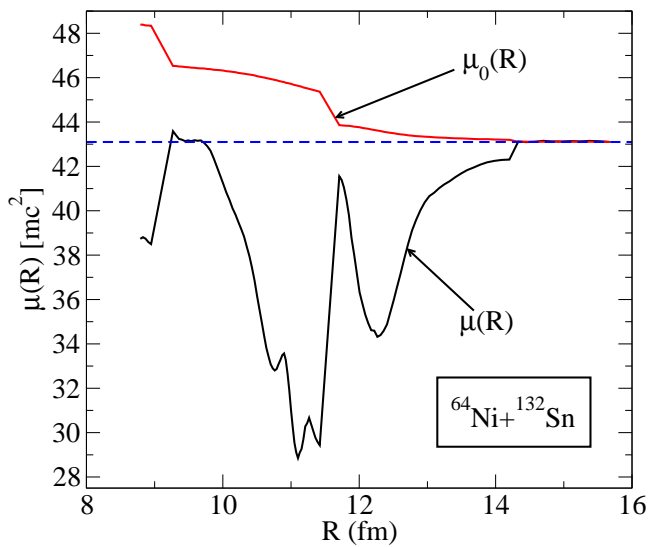}
\caption{\label{fig:mur} (Color online) Reduced mass calculated
for the head-on collision of $^{64}$Ni+$^{132}$Sn system at $E_{\mathrm{c.m.}}=168$~MeV and $\beta=90^{\circ}$.}
\end{figure}

In addition to the effective mass it also possible to calculate the dynamical reduced mass. It is well
known that the naive formula for reduced mass given by
\begin{equation}
\mu _{0}(R)=m\frac{A_{1}(R)A_{2}(R)}{A_{1}(R)+A_{2}(R)}\;,
\end{equation}
does not correctly represent the dynamical behavior of this quantity, monotonically increasing from
its asymptotic value to the predictable final value when $A_1=A_2=(A_1+A_2)/2$, i.e. a composite has
been formed and half the mass is in the right half and other half on the left half of the numerical box.
An alternative way to calculate the dynamical reduced mass is given in Ref.~\cite{DL02}. Here the
dynamical centers and momenta for the left and right halves are calculated via
\begin{eqnarray*}
R_{i}&=&Tr(\hat{{r}}\rho _{i})/Tr(\rho_{i}) \\
P_{i}&=&Tr(\hat{{p}}\rho _{i})\;,
\end{eqnarray*}
where the index $i=1,2$ denotes left and right halves.
The left and right masses are then calculated by
\begin{equation}
m_{i}=\frac{P_{i}}{dR_{i}/dt}\;,
\end{equation}
leading to the dynamical reduced mass
\begin{equation}
\mu (R)=\frac{m_{1}m_{2}}{m_{1}+m_{2}}\;,
\end{equation}
with $R=R_1-R_2$. This quantity is plotted in Fig.~\ref{fig:mur} along with the naive expression, $\mu_0(R)$.
While we observe interesting structure in $\mu(R)$ the overall change in the magnitude around the barrier region
is not large enough to substantially alter the fusion cross sections. This is also shown in Fig.~\ref{fig:eff}
where we have also plotted the ratio of the effective mass to the dynamical reduced mass (dashed curve). Therefore,
in practice we have not included $\mu(R)$ in our cross section calculations.

\subsection{\label{sec:xsection} Cross sections}
\begin{figure}[!htb]
\includegraphics*[scale=0.42]{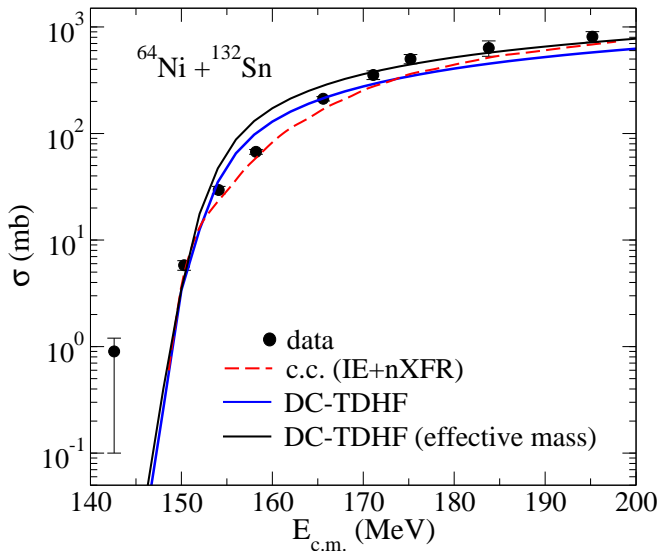}
\caption{\label{fig:xsection} (Color online) Total fusion cross section as a function of $E_{\mathrm{c.m.}}$.
Shown are the experimental data (filled circles), the latest coupled-channel calculation~\cite{Li07}
(red dashed curve), and our density constrained TDHF cross sections with
dynamical effective mass (black curve) and with constant reduced mass (blue curve).}
\end{figure}

We have calculated the total fusion cross section as a function of energy using the alignment averaged
fusion formula given in Eq.~(\ref{eq:fusion}). The total fusion cross section using the potential barriers
obtained from density constrained TDHF calculations directly (with constant reduced mass) is shown by the blue curve
in Fig.~\ref{fig:xsection}. Also shown is the the latest coupled-channel calculation~\cite{Li07}
including inelastic excitation
of $^{64}$Ni to first $2^{+}$ and $3^{-}$ states and $^{132}$Sn to the first $2^{+}$ state, as well as
two-neutron transfer (red dashed curve).
We have also included the dynamical effective mass by making the well known coordinate scale transformation~\cite{GRR83}
\begin{equation}
d\bar{R}=\left(\frac{M(R)}{\mu }\right)^{\frac{1}{2}}dR\;.
\end{equation}
As a result of this transformation all of the effects of the dynamical effective mass are transferred
to the scaled potential while the reduced mass $\mu$ remains constant at its asymptotic value. This is
convenient for a number of reasons, one being that the we can use our fusion code without any modifications.
The resulting cross sections are shown by the black curve in Fig.~\ref{fig:xsection}. As we see the effect
of the dynamical mass is to raise the cross section at higher energies.

With the exception of the lowest energy data point the calculated cross sections are in very good agreement
with data and the extended coupled-channel calculation. We emphasize again that our density constrained TDHF
calculations contain no adjustable parameters or normalization factors. In the region around $160$~MeV our calculations
over predict the experimental cross section while the coupled-channel one slightly under predict.
The agreement at lower and higher energies are excellent. The question about the lowest energy data point
is still an open one. It is our understanding that a new experiment is planned to measure this cross
section~\cite{Li07b}.
If this enhancement is confirmed it would be a challenge since all of the existing
theories underestimate the cross section by a few orders of magnitude. We have shown in our barrier calculations
that this is the only point which is truly sub-barrier fusion. On the other hand such a large value for the cross section
is more consistent with an above the barrier or very close to the top of the barrier energy. This would indicate
that there is a mode of $^{64}$Ni nucleus which is not well described by the current microscopic interactions,
like the Skyrme force. One possibility may be triaxiality since $^{64}$Ni is experimentally found to be gamma soft.

\section{Conclusions}

We have performed density constrained TDHF calculations of fusion cross sections for
the $^{64}$Ni+$^{132}$Sn system. Our results agree very well with the measured data and
with a state of the art
coupled-channel calculation despite having no adjustable parameters. This indicates that many
of the reaction dynamics are included in the TDHF description of the collision.
As we investigate fusion reactions involving neutron rich and deformed nuclei it is
apparent that an understanding of the structure of these nuclei is crucial to the
description of the reaction dynamics. For these nuclei various effects, such as inelastic
excitations, particle transfer, and other dynamical effects lead to substantial modification
of the naive potential barrier calculations which assume an inert core and no dynamics.
Consequently, the definition of {\it sub-barrier} fusion becomes ambiguous since
it is difficult to determine the barrier a priori. Present calculations indicate that
the lowest energy data point may be the only one truly at a sub-barrier energy. If a new
measurement confirms this data point it would be of great interest since none of the
theories can reproduce this cross section, thus indicating a fundamental property that
is not included in any of the calculations.

\begin{acknowledgments}
This work has been supported by the U.S. Department of Energy under grant No.
DE-FG02-96ER40963 with Vanderbilt University.
\end{acknowledgments}




\end{document}